\documentclass[]{aa} 
\usepackage{graphicx}
\usepackage{txfonts}
\usepackage{natbib}
\bibpunct{(}{)}{;}{a}{}{,}
\begin{document}
   \title{Gamma-ray burst detection with the AGILE mini-calorimeter}

   \author{M.~Marisaldi\inst{1}, C.~Labanti\inst{1}, F.~Fuschino\inst{1}, M.~Galli\inst{2}, A.~Argan\inst{3}, G.~Barbiellini\inst{4 , 5}, M.~Basset\inst{5}, F.~Boffelli\inst{6 , 7}, A.~Bulgarelli\inst{1}, P.~Caraveo\inst{8}, P.W.~Cattaneo\inst{6}, A.~Chen\inst{8 , 9}, V.~Cocco\inst{3}, E.~Costa\inst{3}, F.~D'Ammando\inst{3 , 10}, E.~Del~Monte\inst{3}, G.~De~Paris\inst{3}, G.~Di~Cocco\inst{1}, G.~Di~Persio\inst{3}, I.~Donnarumma\inst{3}, Y.~Evangelista\inst{3}, M.~Feroci\inst{3}, A.~Ferrari\inst{9 , 11}, M.~Fiorini\inst{8}, L.~Foggetta\inst{12 , 13}, T.~Froysland\inst{9 , 10}, M.~Frutti\inst{3}, F.~Gianotti\inst{1}, A.~Giuliani\inst{8}, I.~Lapshov\inst{3 , 14}, F.~Lazzarotto\inst{3}, F.~Liello\inst{4 , 5}, P.~Lipari\inst{15 , 16}, F.~Longo\inst{4 , 5}, M.~Mastropietro\inst{17}, E.~Mattaini\inst{8}, A.~Mauri\inst{1}, F.~Mauri\inst{6}, S.~Mereghetti\inst{8}, E.~Morelli\inst{1}, A.~Morselli\inst{18}, L.~Pacciani\inst{3}, A.~Pellizzoni\inst{8}, F.~Perotti\inst{8}, P.~Picozza\inst{18}, C.~Pontoni\inst{5}, G.~Porrovecchio\inst{3}, M.~Prest\inst{12 , 13}, G.~Pucella\inst{3}, M.~Rapisarda\inst{19}, A.~Rappoldi\inst{6}, E.~Rossi\inst{1}, A.~Rubini\inst{3}, P.~Soffitta\inst{3}, M.~Tavani\inst{3 , 10}, A.~Traci\inst{1}, M.~Trifoglio\inst{1}, A.~Trois\inst{3}, E.~Vallazza\inst{5}, S.~Vercellone\inst{8}, V.~Vittorini\inst{10}, A.~Zambra\inst{8 , 9}, D.~Zanello\inst{15 , 16}, C.~Pittori\inst{20}, F.~Verrecchia\inst{20}, S.~Cutini\inst{20}, D.~Gasparrini\inst{20}, B.~Preger\inst{20}, P.~Santolamazza\inst{20}, P.~Giommi\inst{20}, L.A.~Antonelli\inst{20}, S.~Colafrancesco\inst{20}, L.~Salotti\inst{21}
          }
	  
\offprints{M. Marisaldi, \email{marisaldi@iasfbo.inaf.it}}

   \institute{INAF-IASF Bologna, Via Gobetti 101, I-40129 Bologna, Italy
         \and	
         	ENEA, via Martiri di Monte Sole 4, I-40129 Bologna, Italy
        \and	
         	INAF-IASF Roma, via del Fosso del Cavaliere 100, I-00133 Roma, Italy
        \and	
	       	Dipartimento di Fisica Universit\`a di Trieste, via A. Valerio 2, I-34127 Trieste, Italy 
	\and	
		INFN Trieste, via A. Valerio 2, I-34127 Trieste, Italy
        \and	
        	INFN Pavia, via Bassi 6, I-27100 Pavia, Italy
        \and	
		Dipartimento di Fisica Nucleare e Teorica, Universit\`a di Pavia, Via Bassi 6, I-27100 Pavia, Italy
	\and	
          	INAF-IASF Milano, via E. Bassini 15, I-20133 Milano, Italy
        \and	
       		CIFS Torino, Viale Settimio Severo 63, I-10133 Torino, Italy 
         \and	
        	Dipartimento di Fisica, Universit\`a Tor Vergata, via della Ricerca Scientifica 1, I-00133 Roma, Italy
	\and	
        	Dipartimento di Fisica, Universit\`a di Torino, Torino, Italy
       \and	
		Dipartimento di Fisica, Universit\`a dell'Insubria, Via Valleggio 11, I-22100 Como, Italy
 	\and	
		INFN Milano-Bicocca, Piazza della Scienza 3, I-20126 Milano, Italy
        \and	
        	IKI, Moscow, Russia
       \and	
      		INFN Roma ``La Sapienza'', p.le Aldo Moro 2, I-00185 Roma, Italy
        \and	
        	Dipartimento di Fisica, Universit\`a La Sapienza, p.le Aldo Moro 2, I-00185 Roma, Italy
       \and	
        	CNR-IMIP, Area della Ricerca di Montelibretti (Roma), Italy
       \and	
     	  	INFN Roma ``Tor Vergata'', via della Ricerca Scientifica 1, I-00133 Roma, Italy
          \and	
        	ENEA Frascati, via Enrico Fermi 45, I-00044 Frascati(Roma), Italy
         \and	
        	ASI Science Data Center, Via E. Fermi 45, I-00044 Frascati (Roma), Italy
        \and	
        	Agenzia Spaziale Italiana, viale Liegi 26, I-00198 Roma, Italy
  }

   \date{Received ; accepted }

 
  \abstract
   {The Mini-Calorimeter (MCAL) instrument on-board the AGILE satellite is a non-imaging gamma-ray scintillation detector sensitive in the 300~keV--100~MeV energy range with a total on-axis geometrical area of $1400~\mathrm{cm^2}$.
   Gamma-Ray Bursts (GRBs) are one of the main scientific targets of the AGILE mission and the MCAL design as an independent self-triggering detector makes it a valuable all-sky monitor for GRBs. Furthermore MCAL is one of the very few operative instruments with microsecond timing capabilities in the MeV range.}
   {In this paper the results of GRB detections with MCAL after one year of operation in space are presented and discussed.}
  {A flexible trigger logic implemented in the AGILE payload data-handling unit allows the on-board detection of GRBs. For triggered events, energy and timing information are sent to telemetry on a photon-by-photon basis, so that energy and time binning are limited by counting statistics only. When the trigger logic is not active, GRBs can be detected offline in ratemeter data, although with worse energy and time resolution.}
   {Between the end of June 2007 and June 2008 MCAL detected 51 GRBs, with a detection rate of about 1~GRB/week, plus several other events at a few milliseconds timescales. Since February 2008 the on-board trigger logic has been fully active. Comparison of MCAL detected events and data provided by other space instruments confirms the sensitivity and effective area estimations. MCAL also joined the $3^{rd}$ Inter-Planetary Network, to contribute to GRB localization by means of triangulation. }
    {}
\authorrunning{M. Marisaldi et al.}
\titlerunning{GRB detection with the AGILE mini-calorimeter}
   \maketitle
%

\section{Introduction}

The AGILE satellite \citep{Tavani2008,Tavani2008b}, the Italian space mission dedicated to gamma-ray and hard-X astrophysics, has the study of GRBs among its main scientific targets.  The Gamma-Ray Imaging Detector (GRID), composed of a tungsten-silicon tracker \citep{Prest2003} and a CsI(Tl) Mini-Calorimeter, has a wide field of view that makes it a valuable instrument for GRB detection in the poorly explored 30~MeV-50~GeV energy band.
SuperAGILE \citep{Feroci2007}, the hard X-ray imager on-board AGILE operating in the 18-60~keV energy band, is equipped with an on-board trigger logic and localization algorithm providing few arcmin position accuracy, allowing rapid dissemination of the coordinates \citep{DelMonte2007c}. The Mini-Calorimeter, despite being a subsystem of the GRID, is also equipped with a self-triggering operative mode and on-board logic making it an all-sky monitor in the 300~keV-100~MeV energy range.
A simultaneous GRB detection with GRID, MCAL and SuperAGILE would allow spectral coverage over six orders of magnitude. 
In this paper the status of the GRB detection with MCAL, one year after the AGILE launch, is reviewed and discussed.


\section{MCAL GRB detection capabilities}

MCAL is composed of 30 CsI(Tl) scintillator bars (dimensions: $\mathrm{15x23x375~mm^3}$ each) arranged in
two orthogonal layers, for a total thickness of 1.5 radiation lengths. In a
bar the readout of the scintillation light is accomplished
by two custom PIN Photodiodes (PD) coupled one at each short side of the bar. Detailed descriptions of MCAL can be found in \citet{Labanti2006,Labanti2008}.

MCAL works in two possible operative modes:
\begin{itemize}
\item[-] in GRID mode a trigger issued by the silicon tracker starts the collection of all the detector signals in order to determine the energy and position of particles converted in the tracker and interacting in MCAL;
\item[-] in BURST mode each bar behaves as an independent self triggering detector and
generates a continuous stream of gamma-ray events in the energy range 300~keV - 100~MeV. The dynamic range of the electronics was set so high to account also for the high energy emission from GRBs. Nevertheless, due to the limited thickness of the detector, secondaries originated by interactions of photons at energies higher than about 10~MeV are expected to give rise to significant incomplete signal collection, making the energy reconstruction of the incident photon more difficult. 
\end{itemize}
Both operative modes can be active at the same time, but the one relevant to GRB investigation is the BURST mode. BURST data are stored in a circular buffer and analysed by a dedicated trigger logic, described in detail in \citet{Fuschino2008}. If a trigger is issued, the data are sent to telemetry on a photon-by-photon basis including, for each event, energy information and a time tag with $2~\mathrm{\mu s}$ accuracy. Without a trigger, or when the trigger logic is off, due to telemetry limitations BURST data are not sent to the ground on a photon-by-photon basis, but are used to build two broad band energy spectra (Scientific Ratemeters, SRM), one for each detection layer, and stored in telemetry with a 1.024~s time bin. 

Due to programmatic constraints it was not possible to switch on and configure the on-board trigger logic prior to the end of November 2007. Then it was switched off again during January 2008, and since the $5^{th}$ of February 2008 it has again been operative. When the trigger logic was not active, GRBs were detected by on-ground analysis, scanning the SRM data for rate increases with a dedicated software task. Despite several GRBs having been detected with this method, the coarse time and energy binning limits the scientific exploitation of the data. On the contrary, with the onset of the on-board trigger logic, time and energy binning for triggered events is only limited by counting statistics. The early MCAL GRB detections are reported in  \citet{Marisaldi2008}.

Several GRB detectors are currently active in space, each with its own specific characteristics. Apart from Swift-BAT \citep{Barthelmy2000}, INTEGRAL-IBIS \citep{Ubertini2003}, SuperAGILE \citep{Feroci2007}, GLAST-GBM \citep{Meegan2007} and GLAST-LAT \citep{Michelson2007}, all the other detectors have no or very limited imaging capabilities and rely on triangulation between different spacecraft for GRB localization, through the $3^{rd}$ Inter-Planetary Network (IPN)\footnote{IPN web page: http://www.ssl.berkeley.edu/ipn3/}. Among the current IPN instruments, only three have spectroscopic capabilities at MeV energies, in an energy range partially overlapping with that of MCAL: Konus-Wind \citep{Aptekar1995}, Suzaku-WAM \citep{Yamaoka2006} and RHESSI \citep{Wigger2004}. Among these, only the RHESSI spectrometer is capable of photon-by-photon data download. Also GLAST-GBM, expected to join the IPN too, has both spectral capabilities in the MeV range and photon-by-photon data download for triggered events. GLAST-LAT has both high spectral and timing capabilities, but in an energy range higher than that of MCAL.


\section{Results}
\subsection{GRB detections}
Between $22^{nd}$ June 2007 and $30^{th}$ June 2008 MCAL detected 51~GRBs, with an average detection rate of about 1~GRB/week. Most of these detections have been independently confirmed by other instruments. Only 16 events have been localized, either by Swift, SuperAGILE or the IPN, as reported in Table \ref{table:1}. The IPN localizations reported here are those publicly available at the time of writing; since most of the MCAL events have also been detected by other IPN instruments \citep{Hurley2008} the number of IPN localizations is expected to rise when the IPN catalogues become available. The detection rate is in good agreement with the sensitivity estimations reported in \citet{Ghirlanda2004}. It must also be noted that in the same time period SuperAGILE localized another four GRBs that were not detected by MCAL.

%
%
\begin{table}
\begin{minipage}[t]{\columnwidth}
\caption{MCAL GRB detection summary}             
\label{table:1}      
\centering                          
\begin{tabular}{l r}        
\hline\hline                 
MCAL detection & Number of events \\    
\hline                        
 Ground trigger\footnote{Scientific ratemeters only: $22^{nd}$ Jun. 2007 -- $24^{th}$ Nov. 2007, $1^{st}$ Jan. 2008 -- $4^{th}$ Feb. 2008} & 28 \\      
 On-board trigger\footnote{$25^{th}$ Nov. 2007 -- $31^{st}$ Dec. 2007, $5^{th}$ Feb. 2008 -- $30^{th}$ Jun. 2008} & 23 \\
 Localized by SuperAGILE  &  1 \\
 Localized by Swift       & 10 \\
 Localized by IPN \footnote{Public coordinates available in GCN}        &  5 \\
\hline                                   
\end{tabular}
\end{minipage}
\end{table}

\begin{figure*}
   \centering
   \includegraphics[width=7.0in]{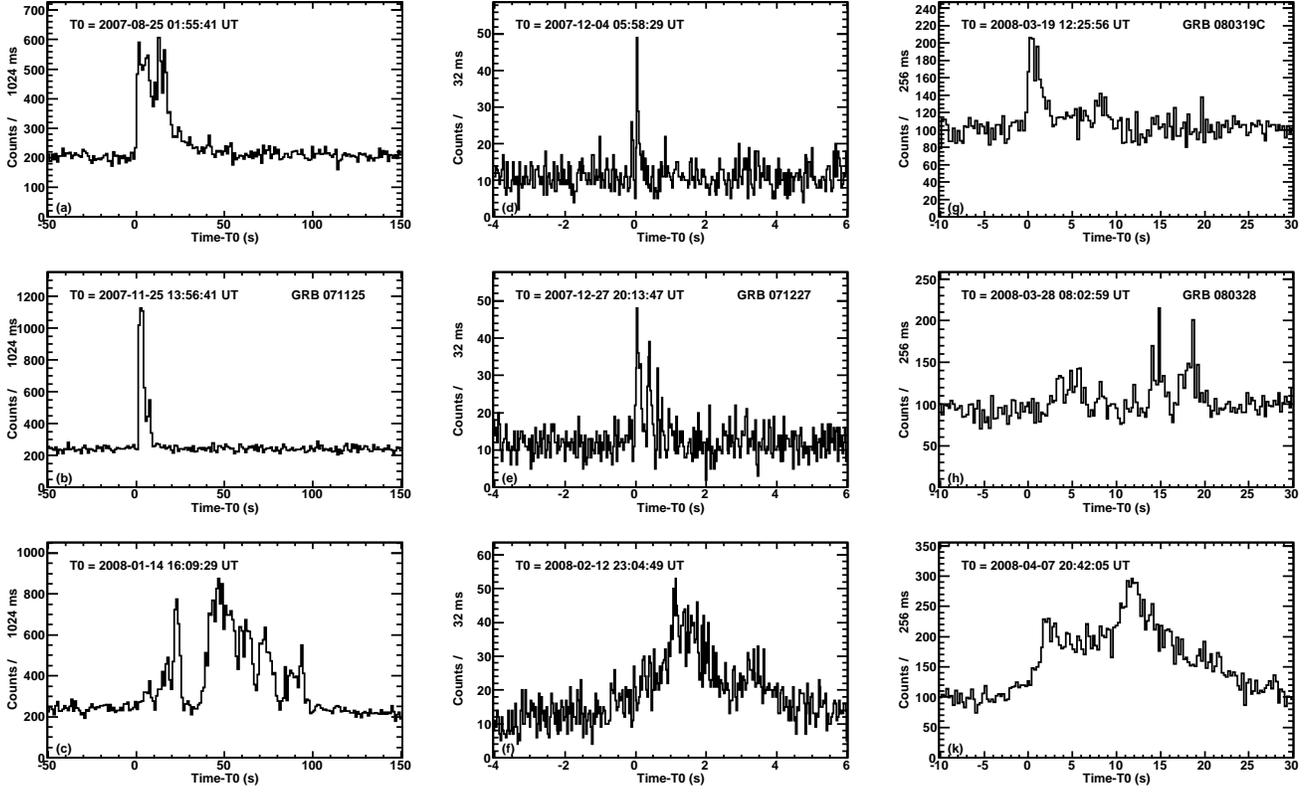}
      \caption{Light curves of several bursts detected by MCAL. Panels (a) to (c): GRBs triggered on the ground by SRM data. Panels (d) to (k): GRBs triggered by the on-board logic.
              }
         \label{MCAL_lc_various}
   \end{figure*}

Figure \ref{MCAL_lc_various} shows the MCAL light curves for a sample of GRBs. Panels (a), (b) and (c) refer to GRBs triggered on the ground based on SRM data. For these events the on-board trigger logic was not active; only the light curves relative to the upper detection layer (the one closer to the silicon tracker) are shown. Panels (d) to (k) refer to GRBs triggered on-board; the light curves relative to the complete instrument are shown. Panels (d) to (f) refer to $<5\mathrm{s}$ long GRBs, shown here with a $32\mathrm{ms}$ time bin. Panels (g) to (k) refer to longer GRBs, shown here with a $256\mathrm{ms}$ time bin. For those events with a public localization available the GRB name is reported too.

%

\begin{figure}
   \centering
   \includegraphics[width=3.5in]{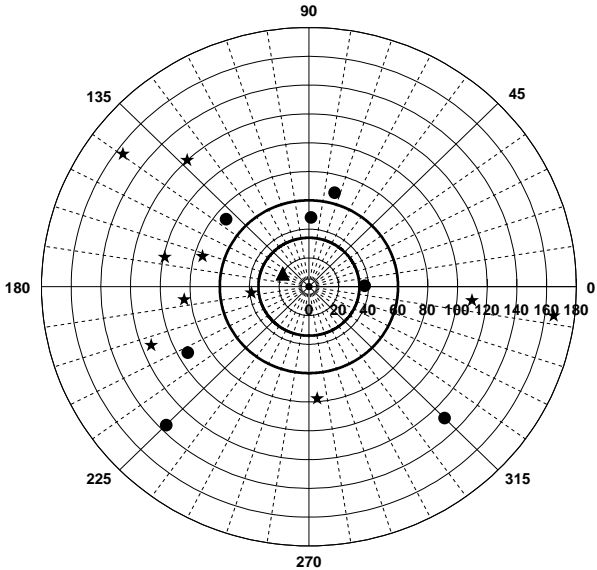}
      \caption{Positions of the GRBs detected by MCAL and localized either by Swift (stars), IPN (circles), or SuperAGILE (triangle) in the reference frame of AGILE pointing. The radial coordinate is the off-axis angle with respect to the pointing of the satellite. The thick solid circles represent the GRID and SuperAGILE (the innermost) fields of view.
}
         \label{FigMCAL_localization_polar}
   \end{figure}

Figure \ref{FigMCAL_localization_polar} is a polar plot centered at the pointing direction of the AGILE satellite, showing the position of the localized GRBs detected by MCAL in the considered period. Coordinates for GRB 070915 and GRB 070825, provided by IPN but not published in GCN, are also included \citep{Palshin2007}. Several GRBs have been detected at off-axis angles greater than $90^\circ$, with the highest being the bright GRB~071020, localized by Swift \citep{Holland2007GCN6949}, detected at $166^\circ$ off-axis, i.e. coming almost from a direction opposite to the AGILE pointing. Despite that for $> 90^\circ$ events it is difficult to provide reliable spectral information, due to the still incomplete modeling of the spacecraft shell with Monte Carlo simulations \citep{Longo2002,Cocco2002b}, it demonstrates the MCAL all-sky detection capabilities.

The only GRB detected by the GRID detector above 50~MeV in the considered period is GRB~080514B \citep{Rapisarda2008,Giuliani2008}, which also triggered MCAL and SuperAGILE and is the subject of a dedicated paper \citep{Giuliani2008b}. This GRB is further considered in subsection \ref{ecal} regarding spectral fitting of MCAL data.

%

\subsection{Sensitivity}
\label{sensitivity}
\begin{figure}
   \centering
   \includegraphics[width=3.5in]{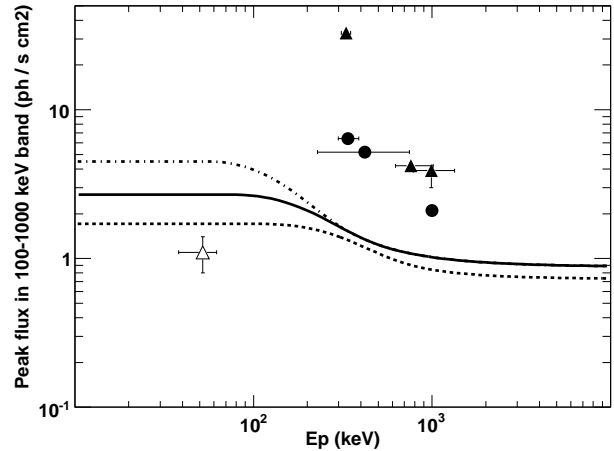}
      \caption{MCAL sensitivity as a function of the peak energy for $40^{\circ}$ off axis GRBs with different Band spectral models. Continuous line: $\alpha=-1$, $\beta=-2.5$; dashed line $\alpha=-0.5$, $\beta=-2$; dot-dashed line $\alpha=-1$, $\beta=-3$. Solid triangles: GRBs detected by MCAL at incident angle $<90^\circ$. Solid circles: GRBs detected by MCAL at incident angle $>90^\circ$. Hollow triangle: GRB at incident angle $<90^\circ$ but not detected by MCAL.
              }
         \label{FigMCAL_sensitivity_theta40}
   \end{figure}

Figure \ref{FigMCAL_sensitivity_theta40} shows the MCAL sensitivity to GRBs. The sensitivity has been calculated according to the procedure described in \citet{Band2003} and reports the peak flux in the $100-1000~\mathrm{keV}$ energy band at the detector's threshold ($5\sigma$ significance above background in 1 second), as a function of the peak energy, for three sample spectral shapes modeled according to the Band Model \citep{Band1993}. The sensitivity curves have been calculated using the MCAL efficiency at a $40^\circ$ off-axis angle resulting from Monte Carlo simulations and the in-flight background level (about 210~counts/s over all the detector in the 330-700~keV trigger band). Superimposed on the sensitivity curves are data points corresponding to GRBs with public spectral parameters, published as GRB Coordinates Network circulars either by the Suzaku-WAM or Konus-Wind teams. Solid triangles refer to GRBs also detected by MCAL at an incident angle lower than $90^\circ$: GRB~071125 \citep{OhnoGCN2007,Golenetskii2007GCN7137}, GRB~080204 \citep{TeradaGCN2008,Golenetskii2008GCN7263}, GRB~080319C \citep{OndaGCN2008,Golenetskii2008GCN7487}. Solid circles refer to GRBs detected by MCAL at incident angles greater than $90^\circ$: GRB~071227 \citep{OndaGCN2008b,GolenetskiiGCN2007}, GRB~080122 \citep{UeharaGCN2008,Golenetskii2008GCN7219}, GRB~080328 \citep{KodakaGCN2008,GolenetskiiGCN2008}. For this plot, the Suzaku-WAM peak flux values have been used because they have been homogeneously provided for the 100~keV--1~MeV band. 

The hollow triangle point below the sensitivity curves corresponds to GRB~071010B, localized by Swift and observed at high energy by Suzaku-WAM \citep{Kira2007GCN6931,Golenetskii2007GCN6879}. This GRB was not detected by MCAL, despite it being unocculted by the Earth and its incoming direction was just $41^\circ$ off-axis, as expected from its spectral parameters and the MCAL sensitivity curve. Other GRBs in similar conditions are GRB~070704 \citep{Kira2007GCN6616,Kira2007GCN6634} and GRB~070724B \citep{Feroci2007GCN6668,Endo2007GCN6672,Golenetskii2007GCN6671}. GRB~070724B has been localized by SuperAGILE and was classified as "No High Energy" \citep{DelMonte2007}.

%

\subsection{Energy calibration}
\label{ecal}

\begin{figure}
   \centering
   \includegraphics[width=3.5in]{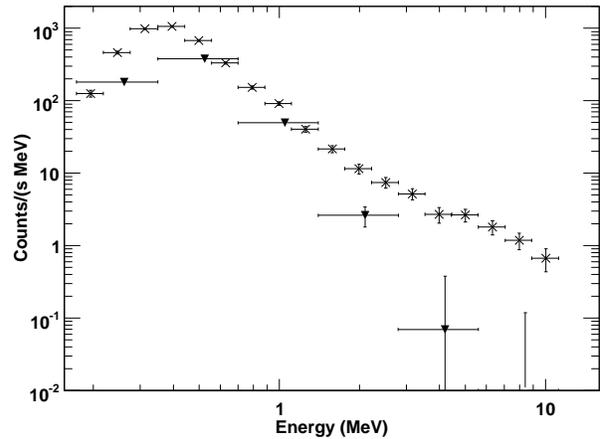}
      \caption{MCAL count spectra for GRB~080407B. Filled triangles: upper detection layer SRM count spectrum. Crosses: count spectrum obtained from photon-by-photon data relative to the whole MCAL. 
              }
         \label{080407_spectra_SRM_TTE}
   \end{figure}

One of the main characteristics of MCAL is the broad energy range extending up to several MeV. To take full advantage of the extended spectral coverage, operation with the on-board logic active is mandatory, as SRMs provide only a coarse spectral resolution. Figure \ref{080407_spectra_SRM_TTE} shows two count spectra for GRB~080407 (trigger time 2008-04-07 20:42:05 UT), the highest fluence GRB triggered on-board by MCAL in the considered period. One spectrum has been obtained from SRM data relative to the upper detection layer (filled triangles), the other is obtained from photon-by-photon data (crosses). Photon-by-photon data obviously allow a much finer spectral reconstruction. Moreover these data are detected in the whole MCAL, while SRM spectra are obtained separately for the two detection layers, making the energy reconstruction more difficult at energies above a few MeV where Compton scattering becomes more important and photons tend to produce multiple hits on different detection layers. The total number of GRB events detected in photon-by-photon mode is about twice that detected with SRM in a single detection layer.

%

\begin{figure}
   \begin{center}
   \begin{tabular}{c}
   \includegraphics[width=3.5in]{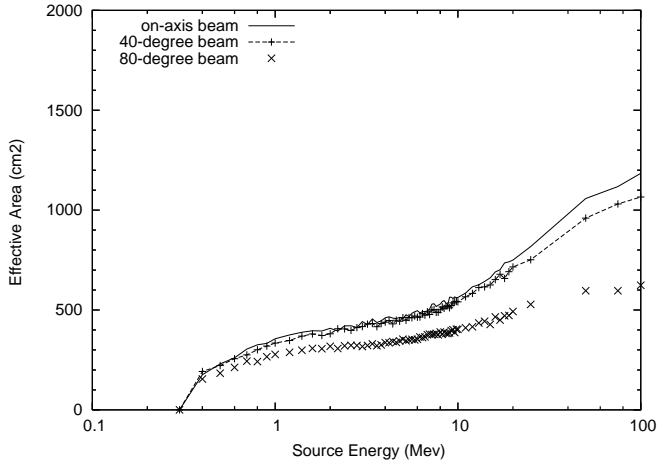}
   \end{tabular}
   \end{center}
   \caption[Aeff] 
   { \label{fig:Aeff} MCAL total effective area as a function of energy, for different incident angles. Solid line: on-axis beam. Pluses: $40^\circ$ off-axis beam. Crosses: $80^\circ$ off-axis beam.
} 
   \end{figure}

Figure \ref{fig:Aeff} shows the MCAL effective area for different off-axis angles calculated from Monte Carlo simulations. The estimated effective area is about $300~\mathrm{cm^2}$ at 1~MeV. Although the effective area increases at higher energies, the limited thickness of the instrument prevents full containment of secondary particles in the pair conversion regime. At low energy, between 330~keV and about 1~MeV, the effective area is strongly dependent on the energy threshold of each detector's bar. Moreover the error on energy estimation, based on weighing the signals from both photodiodes for each bar, becomes larger as the energy approaches the threshold, so the low energy response requires careful calibration. Details on the MCAL effective area and energy estimation algorithm are reported in \citet{Labanti2008}. 

\begin{figure}
   \centering
   \includegraphics[height=3.3in, angle=-90]{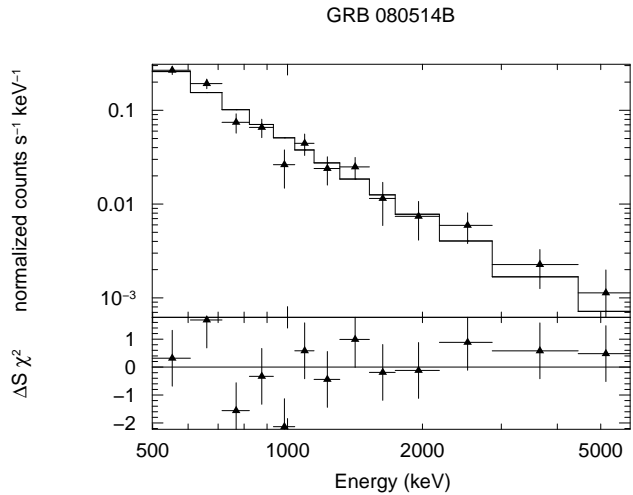}
      \caption{Spectral fitting of GRB 080514B. Filled triangles: MCAL data. Solid line: best fit power law model.}
         \label{080514B_MCAL}
   \end{figure}

Figure \ref{080514B_MCAL} shows the spectral fitting with a power law model for the MCAL data of GRB~080514B, in the energy range 500--5000~keV, using preliminary calibration parameters. The solid line represents the best-fit model folded with the MCAL response (reduced $\chi^2 = 1.1 $ with 12 degrees of freedom). The photon index is $-2.66^{+0.30}_{-0.25}$, where quoted errors are at the 90\% confidence level. This value is consistent within errors with the high energy photon index obtained fitting Konus-Wind and Suzaku-WAM data with a Band model, as reported in \citet{Golenetskii2008GCN7751} and \citet{Hanabata2008GCN7752}. As observed by MCAL, the burst fluence in the 500--5000~keV energy range is $(7.8 \pm 1.5) \cdot 10^{-6}~\mathrm{erg/cm^2}$ which is about 60\% of the value expected integrating the Konus-Wind spectra in the same energy range. Below 500~keV the number of observed events is lower than expectations and an acceptable fit cannot be obtained, at the moment. This could be due to an overestimation of the effective area at low energy. To validate the effective area using in-flight data, a cross calibration activity is currently ongoing using bright GRBs detected both by MCAL and Konus-Wind.  The low energy effective area assessment, in order to properly estimate the peak energy for detected bursts, as well as the total fluence discrepancy are the most important items to be addressed by the cross calibration activity. Due to the fact that the on-board trigger logic has been steadily active since February 2008, at the time of writing only a few GRB events are suitable for this activity, which is currently in progress.


%

\subsection{Timing performance}

   \begin{figure}
   \centering
   \includegraphics[width=3.5in]{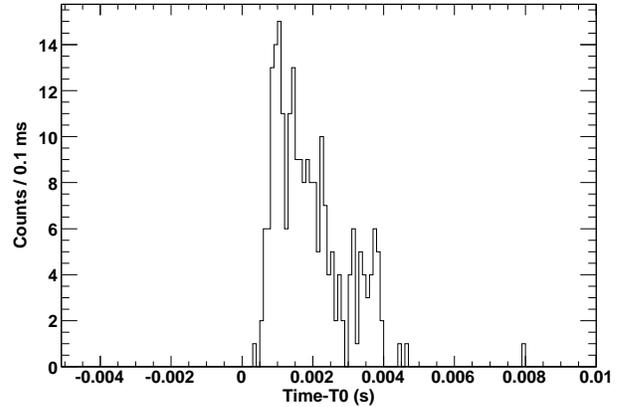}
      \caption{Light curve of the event triggered on 2008-03-23 at 02:28:25 UT
              }
         \label{Fig080323}
   \end{figure}
%

The MCAL high time resolution capability is one of the remarkable characteristics of the instrument, and can be clearly inferred from the light curve reported in Fig.~\ref{Fig080323}. This event was triggered on-board on 2008-03-23 at 02:28:25~UT in the $64~\mathrm{ms}$ time window. It has a duration of about $4~\mathrm{ms}$ and it is bright enough to be binned with $100~\mathrm{\mu s}$ time bins. Since the activation of the on-board trigger logic, several events like this one have been detected \citep{Fuschino2008b}. 
Despite the terrestrial origin of these gamma-ray flashes being considered the most probable explanation \citep{Fishman1994,Smith2005}, their possible cosmic origin is under investigation as well. 


\section{Conclusions}

MCAL is tailored to the detection of medium-bright GRBs with peak energies above a few hundred keV, as expected from sensitivity calculations and experimental evidence reported in section \ref{sensitivity}. The main characteristics of the instrument are its spectroscopic capabilities in the MeV range and the microsecond timing accuracy. Between $22^{nd}$ June 2007 and $30^{th}$ June 2008 MCAL detected 51 GRBs, with an average detection rate of about 1~GRB/week. A preliminary flux calibration is in good agreement with expectations. Since the beginning of February 2008 the on-board trigger logic has been active and calibration activities are in progress.

\begin{acknowledgements}
      We wish to thank Kevin Hurley and Valentin Pal'shin for fruitful discussion and IPN support.
\end{acknowledgements}

\bibliographystyle{aa}
\bibliography{./0562.bib}
 
\end{document}